\documentclass[sigconf,nonacm]{acmart}
\usepackage{subcaption}
\usepackage{amsmath}
\usepackage{xcolor} 
\AtBeginDocument{%
  }

\setcopyright{acmlicensed}
\copyrightyear{2018}
\acmYear{2018}
\acmDOI{XXXXXXX.XXXXXXX}
\acmConference[Conference acronym 'XX]{Make sure to enter the correct
  conference title from your rights confirmation email}{June 03--05,
  2018}{Woodstock, NY}
\acmISBN{978-1-4503-XXXX-X/2018/06}

\begin{document}

\title{Towards Fast Option Pricing PDE Solvers Powered by PIELM}

\author{Akshay Govind Srinivasan}
\authornote{All authors contributed equally to this research.}
\email{me22b102@smail.iitm.ac.in}
\orcid{0009-0001-8740-8281}
\author{Anuj Jagannath Said}
\authornotemark[1]
\email{me21b172@smail.iitm.ac.in}
\orcid{0009-0000-7484-233X}
\author{Sathwik Pentela}
\authornotemark[1]
\email{da24m017@smail.iitm.ac.in}
\orcid{0009-0006-5849-3284}
\affiliation{%
  \institution{Indian Institute of Technology, Madras}
  \city{Chennai}
  \state{Tamil Nadu}
  \country{India}
}

\author{Vikas Dwivedi}
\authornotemark[2]
\orcid{0000-0003-3681-9168}
\email{vikas.dwivedi@creatis.insa-lyon.fr}
\affiliation{
  \institution{Creatis Biomedical Imaging Laboratory, INSA-Lyon,}
  \country{France}}
\author{ Balaji Srinivasan }
\authornotemark[1]
\orcid{0000-0002-3118-5779}
\email{ sbalaji@dsai.iitm.ac.in}
\affiliation{%
  \institution{Indian Institute of Technology, Madras}
  \city{Chennai}
  \state{Tamil Nadu}
  \country{India}
}
\begin{abstract}
Partial differential equation (PDE) solvers underpin modern quantitative finance, governing option pricing and risk evaluation. Physics-Informed Neural Networks (PINNs) have emerged as a promising approach for solving the forward and inverse problems of partial differential equations (PDEs) using deep learning. However they remain computationally expensive due to their iterative gradient descent based optimization and scale poorly with increasing model size. This paper introduces \emph{Physics-Informed Extreme Learning Machines} (PIELMs) as  fast alternative to PINNs for solving both forward and inverse problems in financial PDEs. PIELMs replace iterative optimization with a single least-squares solve, enabling deterministic and efficient training. We benchmark PIELM on the Black–Scholes and Heston–Hull–White models for forward pricing and demonstrate its capability in inverse model calibration to recover volatility and interest rate parameters from noisy data. From experiments we observe that PIELM achieve accuracy comparable to PINNs while being up to $30\times$ faster, highlighting their potential for real-time financial modeling. For reproducibility, codes are available at: \url{https://anonymous.4open.science/r/PIELM-for-Option-Pricing-66CB}
\end{abstract}


  \label{fig:teaser}

\received{20 February 2007}
\received[revised]{12 March 2009}
\received[accepted]{5 June 2009}

\maketitle

\section{Introduction}
\label{sec:intro}

Partial Differential Equations (PDEs) and Ordinary Differential Equations (ODEs) form the backbone of modern quantitative finance, governing the evolution of derivative prices, risk measures, and interest rates. 
For example, the celebrated Black--Scholes framework yields a parabolic PDE for European option prices under the risk-neutral measure~\cite{black1973pricing}. 
More advanced models such as the Heston model for stochastic volatility or the Hull--White model for stochastic interest rates lead to higher-dimensional PDEs that lack closed-form solutions~\cite{heston1993closed,hull1990pricing}. 

Mesh-free approaches using neural networks have become a popular alternative~\cite{BANSAL20251,hoshisashi2024physics,gao2025adaptivemovementsamplingphysicsinformed,sungwon2025asian}. 
Physics-Informed Neural Networks (PINNs)~\cite{raissi2019physics} encode PDEs and boundary conditions into the training loss of a neural network, enabling solution of both forward and inverse problems in finance~\cite{salvador2020financial}. 
PINNs have been applied successfully to a variety of financial models, including stochastic volatility and interest rate frameworks, as well as to parameter inference tasks. 
However, their reliance on iterative gradient-based optimization over large networks results in long runtimes, often orders of magnitude slower than classical solvers for comparable accuracy~\cite{McGreivy2024}. 
Furthermore, PINNs require careful hyperparameter tuning and their hidden-layer representations lack interpretability, making deployment in time-sensitive financial applications challenging~\cite{krishnapriyan2021characterizing}.

Extreme Learning Machines (ELMs)~\cite{huang2006extreme} provide an efficient alternative. 
ELMs employ a single hidden layer with randomly initialized input weights, while output weights are obtained in closed form via a pseudo-inverse solve. 
This eliminates iterative backpropagation, reducing training to a one-shot linear regression while preserving universal approximation guarantees~\cite{huang2006universal}. 
Building on this, Physics-Informed Extreme Learning Machines (PIELMs)~\cite{dwivedi2019physics} integrate PDE and boundary constraints into the ELM framework, forming a linear system from PDE residuals and observed data. 
PIELMs have been shown to achieve comparable accuracy to PINNs, while training several orders of magnitude faster~\cite{dwivedi2020physics,xu2022bayesian}.

The computational advantage of PIELMs is significant. 
Unlike PINNs, which require many iterations of gradient descent over large networks, PIELMs solve a single linear system to obtain the network coefficients, reducing training time from minutes or hours to seconds in many cases. 
This deterministic, one-shot training makes PIELMs particularly suitable for real-time or high-frequency financial applications, where rapid evaluation of option prices or model calibration is essential~\cite{dwivedi2020physics,xu2022bayesian}.

\paragraph{Contributions.} 
In this work, we investigate PIELMs as fast, mesh-free solvers for financial PDEs. Specifically:
\begin{itemize}
    \item We extend PIELMs to both forward and inverse problems in option pricing, focusing on European options under the Black--Scholes model and the Heston--Hull--White stochastic volatility and interest rate model.
    \item We demonstrate that PIELMs achieve high accuracy while drastically reducing computation time, making them viable for real-time applications.
    \item We present a Bayesian framework for inverse problems using PIELMs, allowing efficient recovery of model parameters from noisy observations while enforcing PDE consistency.
\end{itemize}

The remainder of this paper is organized as follows: 
Section~\ref{sec:rbf-pielm-math} presents the mathematical formulation of PIELM; 
Section~\ref{sec:exp} introduces the benchmark problems and experimental setup; 
Section~\ref{sec:results} reports results for both forward and inverse problems; 
and Section~\ref{sec:conclusion} concludes with discussion and future directions.

\section{Methodology}
\label{sec:rbf-pielm-math}

\subsection{Forward Problem Formulation}
\label{sec:forward-formulation}

Let $u:\Omega\subset\mathbb{R}^m\to\mathbb{R}^n$ be a function that satisfies the general linear differential operator equation
\begin{equation}
\label{eqn:linear-op}
    \mathcal{L}(u)(x) + f(x) = 0, \quad x \in \Omega,
\end{equation}
subject to boundary conditions
\begin{equation}
    \mathcal{B}(u)(x) = g(x), \quad x \in \partial\Omega,
\end{equation}
where $\mathcal{L}$ and $\mathcal{B}$ denote the differential and boundary operators, respectively, and $f(x)$ is a known source term. 

To approximate the solution $u(x)$, we employ a single-hidden-layer neural network, known as the Physics-Informed Extreme Learning Machine (PI-ELM). 
The hidden layer parameters (weights and biases) are randomly initialized and fixed, while only the output weights are determined analytically. 
We postulate the functional form
\begin{equation}
\label{eqn:fnapprox}
    \hat{u}(x) = \sum_{i=1}^{N^*} c_i\,\phi_i(x),
\end{equation}
where $\phi_i(x)$ are nonlinear activation functions and $N^*$ denotes the number of hidden neurons. 
In this work, we adopt the hyperbolic tangent activation:
\begin{equation}
\label{eqn:tanh-act}
    \phi(x) = \tanh(mx + b),
\end{equation}
where $m$ and $b$ are random vectors representing the hidden layer weights and biases, respectively. 

We define interior collocation points $\{x_j^{\Omega}\}_{j=1}^{N_\Omega}$ and boundary collocation points $\{x_k^{\partial}\}_{k=1}^{N_{\partial}}$. 
The corresponding residuals for the governing PDE and boundary conditions are given by
\begin{align}
    \mathcal{R}_{\Omega}(x) &= \mathcal{L}(\hat{u})(x) + f(x), \quad x \in \Omega, \\
    \mathcal{R}_{\partial}(x) &= \mathcal{B}(\hat{u})(x) - g(x), \quad x \in \partial\Omega.
\end{align}
Enforcing these residuals at collocation points yields an overdetermined linear system, assuming that the total number of collocation points exceeds the number of neurons ($N_\Omega + N_\partial > N^*$):
\begin{equation}
\label{eqn:linear-system}
    A\,c = b,
\end{equation}
where $A$ collects the PDE and boundary residuals and $c = [c_1,\dots,c_{N^*}]^\top$ are the trainable output weights. 
The least-squares solution is obtained via the Moore–Penrose pseudoinverse or a regularized linear solver:
\begin{equation}
    c = A^{\dagger}b.
\end{equation}
Substituting $c$ back into Equation~\eqref{eqn:fnapprox} provides the physics-informed approximation $\hat{u}(x)$ satisfying the governing equations and boundary constraints in a least-squares sense. 
Unlike standard neural networks, PI-ELMs avoid iterative backpropagation and allow an explicit, closed-form solution for $c$, resulting in extremely fast convergence while preserving the physics constraints.

\vspace{0.5em}
\noindent\textbf{Application to the Black--Scholes Equation.}
For the Black--Scholes partial differential equation
\begin{equation}
    \frac{\partial V}{\partial t}
    + \frac{1}{2}\sigma^2 S^2 \frac{\partial^2 V}{\partial S^2}
    + r S \frac{\partial V}{\partial S}
    - r V = 0,
\end{equation}
the operator $\mathcal{L}$ acts on the neural approximation $\hat{V}(S,t)$ through automatic differentiation of the tanh activations. 
The PI-ELM solution satisfies prescribed terminal and boundary conditions, providing a smooth and differentiable surrogate for the option price $V(S,t)$.

\subsection{Inverse Problem Formulation}
\label{sec:inverse-formulation}

While the forward problem aims to approximate the field $u(x)$ for known operator coefficients, the \textit{inverse problem} seeks to identify unknown physical parameters within the operator $\mathcal{L}_{\boldsymbol{\theta}}$ from available data. 
Let $\boldsymbol{\theta} = [\theta_1, \theta_2, \dots, \theta_p]^\top$ denote the set of unknown parameters, and define the parametric PDE as
\begin{equation}
\label{eqn:parametric-pde}
    \mathcal{L}_{\boldsymbol{\theta}}(u)(x) + f(x) = 0, \quad x \in \Omega.
\end{equation}
Given a dataset of noisy or partial measurements
$\mathcal{D} = \{(x_i, u_i^{\text{obs}})\}_{i=1}^{N_d}$,
the goal is to infer $\boldsymbol{\theta}$ such that the PI-ELM prediction $\hat{u}(x; \boldsymbol{\theta})$ best matches the observed data.

The inverse problem can be cast as an optimization task:
\begin{equation}
\label{eqn:inverse-opt}
    \boldsymbol{\theta}^* = 
    \arg\min_{\boldsymbol{\theta} \in \Theta}
    \frac{1}{N_d}\sum_{i=1}^{N_d}
    \big|\hat{u}(x_i;\boldsymbol{\theta}) - u_i^{\text{obs}}\big|^2,
\end{equation}
where $\Theta$ denotes the admissible parameter space and the loss function measures the mean-squared discrepancy between predicted and observed states.

In this work, we adopt a \textbf{Bayesian Optimization (BO)} strategy to efficiently search the parameter space $\Theta$ without requiring gradients of the loss function. 
At each BO iteration $k$, a trial parameter set $\boldsymbol{\theta}^{(k)}$ is sampled from a probabilistic surrogate model (Gaussian Process or Tree-structured Parzen Estimator), and the forward PI-ELM solver is trained using the operator $\mathcal{L}_{\boldsymbol{\theta}^{(k)}}$. 
The resulting prediction $\hat{u}(x;\boldsymbol{\theta}^{(k)})$ is evaluated against the observed data to compute the misfit $\mathcal{J}(\boldsymbol{\theta}^{(k)})$, which is then used to update the surrogate model. 
This iterative process continues until convergence or until a predefined number of trials is reached.

\vspace{0.5em}
\noindent\textbf{Inverse Black--Scholes Problem.}
For the Black--Scholes equation, the parameters of interest are the volatility $\sigma$ and the risk-free rate $r$, collectively denoted by $\boldsymbol{\theta} = [\sigma, r]^\top$. 
Given a set of observed option prices $V^{\text{obs}}(S_i,t_i)$, the inverse problem is defined as
\begin{equation}
\label{eqn:bs-inverse}
    (\sigma^*, r^*) = 
    \arg\min_{\sigma \in [0,1],\; r \in [0,0.1]}
    \frac{1}{N_d}\sum_{i=1}^{N_d}
    \big|\hat{V}(S_i,t_i;\sigma,r) - V_i^{\text{obs}}\big|^2.
\end{equation}
For each candidate pair $(\sigma, r)$ proposed by the Bayesian optimizer, the PI-ELM forward solver reconstructs the option price field $\hat{V}(S,t;\sigma,r)$ and computes the associated data misfit. 
The BO algorithm iteratively refines the parameter estimates by balancing exploration and exploitation in the search space, yielding the optimal $(\sigma^*, r^*)$ that minimize the prediction error.

This hybrid methodology seamlessly couples the physics-based structure of PI-ELMs with the global search capability of Bayesian Optimization, enabling robust and data-efficient parameter inference for complex PDE systems.


\section{Numerical Experiments}
\label{sec:exp}
To evaluate the accuracy and efficiency of our proposed method, we carried out a set of numerical experiments on benchmark problems from option pricing. Two representative forward cases were selected from the literature: (i) a standard European call option under the Black–Scholes model~\cite{salvador2020financial}, and (ii) a European call option under the Heston–Hull–White framework~\cite{heston1993closed}. 

Additionally, we tested the PIELM framework on an inverse problem, where the objective is to recover model parameters such as volatility and risk-free rate directly from noisy observations of option prices.

\subsection{Case 1: Black--Scholes European Call Option}
The first benchmark follows \cite{salvador2020financial}, where the European call option is priced under the classical Black–Scholes PDE:
\begin{equation}
\frac{\partial V}{\partial t} 
+ \frac{1}{2}\sigma^2 S^2 \frac{\partial^2 V}{\partial S^2} 
+ rS \frac{\partial V}{\partial S} 
- rV = 0, 
\label{eq:bs-pde}
\end{equation}
subject to the terminal condition
\[
V(T,S) = \max(S-K,0).
\]
The boundary conditions are imposed as:
\begin{align*}
V(t,0) &= 0, &\quad & S \to 0, \\
\frac{\partial^2 V}{\partial S^2} &\to 0, &\quad & S \to \infty.
\end{align*}

\paragraph{Parameters.}  
We use the setup in \cite{salvador2020financial}, summarized in Table~\ref{tab:bs-params}.
\begin{table}[H]
\centering
\caption{Black–Scholes European Call Option Parameters}
\label{tab:bs-params}
\begin{tabular}{lc}
\toprule
Strike price ($K$) & $1.0$ \\
Maturity ($T$) & $1.0$ year \\
Risk-free rate ($r$) & $5.0\%$ p.a. \\
Volatility ($\sigma$) & $20\%$ p.a. \\
\bottomrule
\end{tabular}
\end{table}

\begin{figure*} 
\centering \includegraphics[width=1\linewidth]{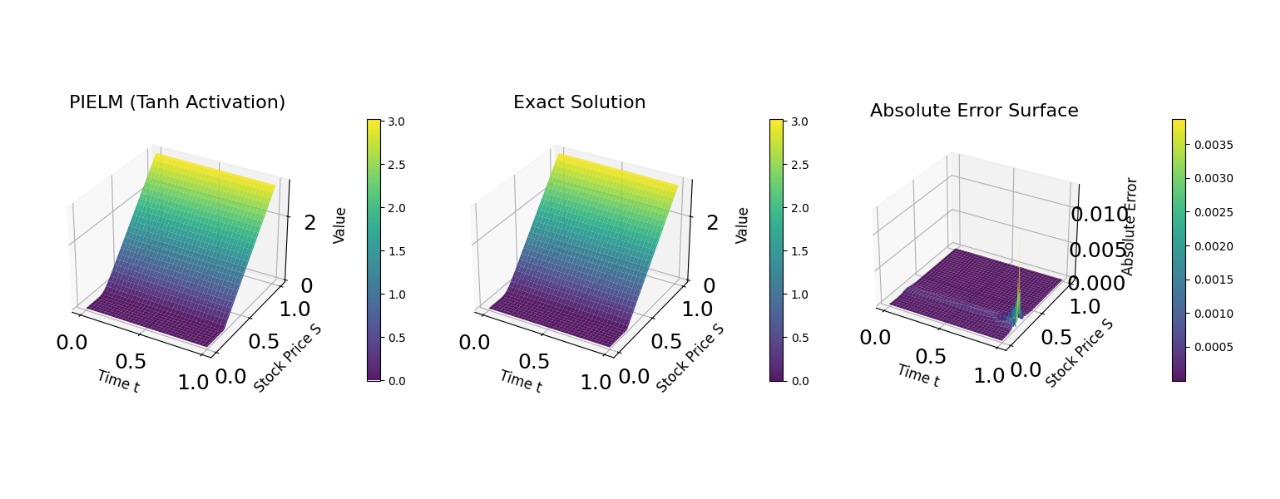} \caption{Solution Obtained for European Call Option using PIELM.} \label{fig:tanh-euro-call} 
\end{figure*}  

\subsection{Case 2: Heston--Hull--White European Call Option}

The second benchmark follows the Heston–Hull–White (HHW) framework, which extends the Black–Scholes model by allowing both stochastic volatility and stochastic short rates \cite{heston1993closed}. The coupled risk-neutral dynamics are given by:
\begin{align}
dS_t &= r_t S_t dt + \sqrt{v_t} S_t dW_t^S, \notag \\
dv_t &= \kappa_v(\theta_v - v_t)dt + \sigma_v \sqrt{v_t} dW_t^v, \notag \\
dr_t &= \kappa_r(\theta_r - r_t)dt + \sigma_r dW_t^r, \notag
\end{align}
with correlations 
\[
dW_t^S dW_t^v = \rho_{Sv} dt, \quad 
dW_t^S dW_t^r = \rho_{Sr} dt, \quad 
dW_t^v dW_t^r = \rho_{vr} dt.
\]

The corresponding PDE for the option value $V(t,S,v,r)$ is:
\begin{align}
\frac{\partial V}{\partial t} 
+ rS \frac{\partial V}{\partial S} 
+ \kappa_v(\theta_v - v)\frac{\partial V}{\partial v} 
+ \kappa_r(\theta_r - r)\frac{\partial V}{\partial r} 
+ \tfrac{1}{2} v S^2 \frac{\partial^2 V}{\partial S^2} \notag \\
+ \rho_{Sv}\sigma_v vS \frac{\partial^2 V}{\partial S \partial v} 
+ \rho_{Sr}\sigma_r S \sqrt{v} \frac{\partial^2 V}{\partial S \partial r} 
+ \tfrac{1}{2}\sigma_v^2 v \frac{\partial^2 V}{\partial v^2} 
+ \tfrac{1}{2}\sigma_r^2 \frac{\partial^2 V}{\partial r^2} 
- rV = 0,
\label{eq:hhw-pde}
\end{align}
with terminal condition $V(T,S,v,r) = \max(S-K,0)$. The boundary conditions are chosen to reflect vanishing option value as $S\to0$, linear growth as $S\to\infty$, deterministic forward payoff when $v\to0$, and sensitivity saturation when $v\to\infty$.

\begin{table}[H] \centering \caption{Heston–Hull–White European Call Option Parameters} 
\label{tab:hhw-params} 
\begin{tabular}{lc} \toprule Strike price ($K$) & $1.0$ \\ Maturity ($T$) & $2.0$ years \\ Initial short rate ($r_0$) & $5.0\%$ p.a. \\ Hull–White mean reversion ($\kappa_r$) & $0.2$ \\ Hull–White long-term mean ($\theta_r$) & $3.0\%$ \\ Hull–White volatility ($\sigma_r$) & $1.0\%$ p.a. \\ Heston initial variance ($v_0$) & $0.5$ \\ Heston mean reversion ($\kappa_v$) & $2.0$ \\ Heston long-term variance ($\theta_v$) & $14\%$ p.a. \\ Heston vol-of-vol ($\sigma_v$) & $50\%$ p.a. \\ Correlations $(\rho_{Sv}, \rho_{Sr}, \rho_{vr})$ & $(-0.6, 0.1, 0.0)$ \\ \bottomrule \end{tabular} \end{table}

\begin{figure}[htbp] \centering \includegraphics[width=\linewidth]{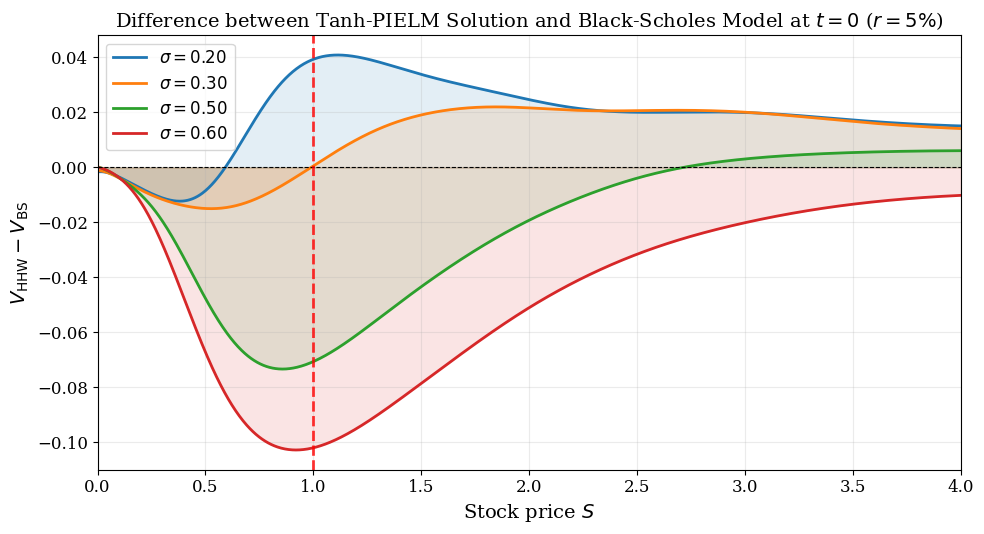} \caption{Difference between Heston-Hull-White (calculated using PIELM) and Black-Scholes call prices as a function of volatility. Mean reversion in the Heston model produces the characteristic bell shape.} \label{fig:stoc-r5} \end{figure} 

\begin{figure}[htbp] \centering \includegraphics[width=\linewidth]{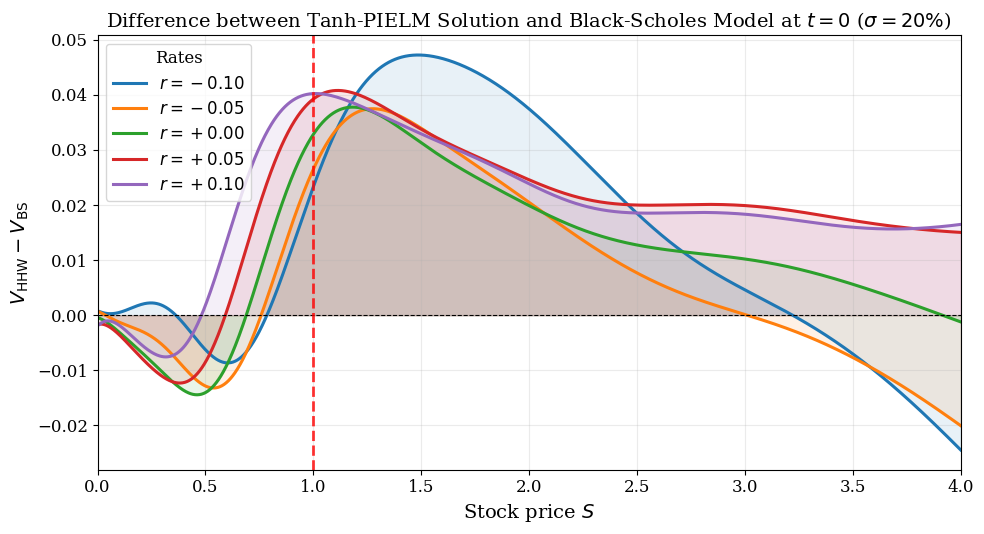} \caption{Difference between Heston-Hull-White and Black-Scholes call prices as a function of the short rate. Stochastic mean reversion of rates leads to systematically higher prices than under constant rate discounting, particularly when initial rates are low.} \label{fig:stoc-sigma20} \end{figure}

\begin{figure*} \centering \includegraphics[width=1\linewidth]{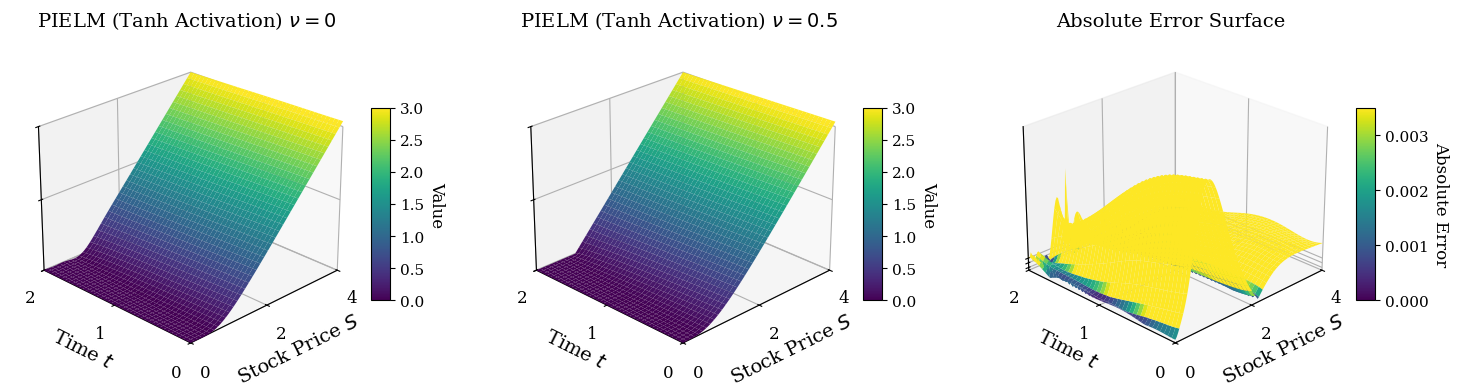} \caption{European Call Option with stochastic volatility and stochastic rates PIELM solution.} \label{fig:euro-call-stoc} \end{figure*} 
\subsection{Case 3: Inverse Problem}
\label{sec:inverse-experiments}

To validate the proposed Bayesian PIELM inverse formulation, we conduct experiments using synthetically generated option price data under known volatility and risk-free rate parameters using~\ref{eq:bs-pde}. The trained forward PIELM serves as the PDE solver, while the inverse model iteratively updates the parameters $\theta = [\sigma, r]$ to minimize the loss $\mathcal{L}_{\text{inv}}$.

\section{Results}
\label{sec:results}

\begin{table*}
\centering
\caption{Error metrics for European Call Option pricing using PIELM and PINN methods.}
\label{tab:results}
\begin{tabular}{lcccc}
\toprule
\textbf{Method} & \textbf{MSE} & \textbf{Relative $\ell_2$} & \textbf{Run time (s)} & \textbf{Rel. Runtime} \\
\midrule
\multicolumn{5}{l}{\textbf{European Call Option (Black-Scholes)}} \\
\cmidrule(lr){1-5}
PINN (baseline) & $4.19 \times 10^{-5}$ & $4.05 \times 10^{-9}$ & $108$ & $1\times$ \\
$tanh$-PIELM & $6.97 \times 10^{-8}$ & $7.5 \times 10^{-9}$ & $18.37$ & $5.89\times$ \\
\midrule
\multicolumn{5}{l}{\textbf{European Call Option (Heston-Hull-White)}} \\
\cmidrule(lr){1-5}
PINN (baseline) & $4.98 \times 10^{-5}$ & $4.61 \times 10^{-3}$ & 6120 & $1\times$ \\
$tanh$-PIELM & $7.66 \times 10^{-3}$ & $9.06 \times 10^{-5}$ & 225.1 & $27.19\times$ \\
\bottomrule
\end{tabular}
\end{table*} 
\subsection{Case 1: Black--Scholes European Call Option}
From Figure ~\ref{fig:tanh-euro-call},  we observe that PIELM closely reproduce the closed-form solution of the Black--Scholes PDE. In terms of computational performance, PIELM achieves solutions $6.14\times$ faster than the reference implementation, while PIELM achieves $5.89\times$ faster, with similar MSE and relative $l_2$ error as compared to our baseline PINN. The error peaks near the strike price, which is expected due to the high gradient in the payoff function.

\subsection{Case 2: Heston--Hull--White European Call Option}
Figure~\ref{fig:euro-call-stoc} depicts the option price surface generated under the Heston-Hull-White (HHW) model using the proposed PIELM approach. The surface exhibits the expected nonlinear behavior in both volatility and interest rate dimensions, similar to what is observed in simpler models like Black-Scholes. To obtain a qualitative benchmark, we compare PIELM results with the Black-Scholes model in Figures~\ref{fig:stoc-r5} and ~\ref{fig:stoc-sigma20}. 

Figure~\ref{fig:stoc-r5} shows the difference in call prices between the HHW and Black-Scholes models as a function of initial volatility $\sigma$. The bell-shaped profiles arise due to the mean-reverting nature of the stochastic volatility in the Heston component: very low or very high initial volatilities tend to revert to the long-term mean, reducing their effect over the option’s life. The largest differences occur near the strike, where vega—the sensitivity of option price to volatility—is highest. In contrast, deep in-the-money or out-of-the-money options are less affected, resulting in flatter regions of the curve. Positive differences indicate scenarios where the stochastic volatility reverts to a level higher than the constant volatility assumed by Black-Scholes, and vice versa. 

Similarly, Figure~\ref{fig:stoc-sigma20} illustrates the effect of stochastic interest rates. When the initial rate is low, mean reversion leads to reduced discounting compared to the constant rate in Black-Scholes, increasing the option price. Conversely, when the rate is high, mean reversion dampens the discounting effect, narrowing the price gap. Overall, our method is able to capture the combined effect of stochastic volatility and interest rates efficiently and produces results $32.17 \times$ times faster than PINNs, while maintaining similar order of MSE.

\subsection{Case 3: Inverse Problem }
\begin{figure}
\centering
\includegraphics[width=\linewidth]{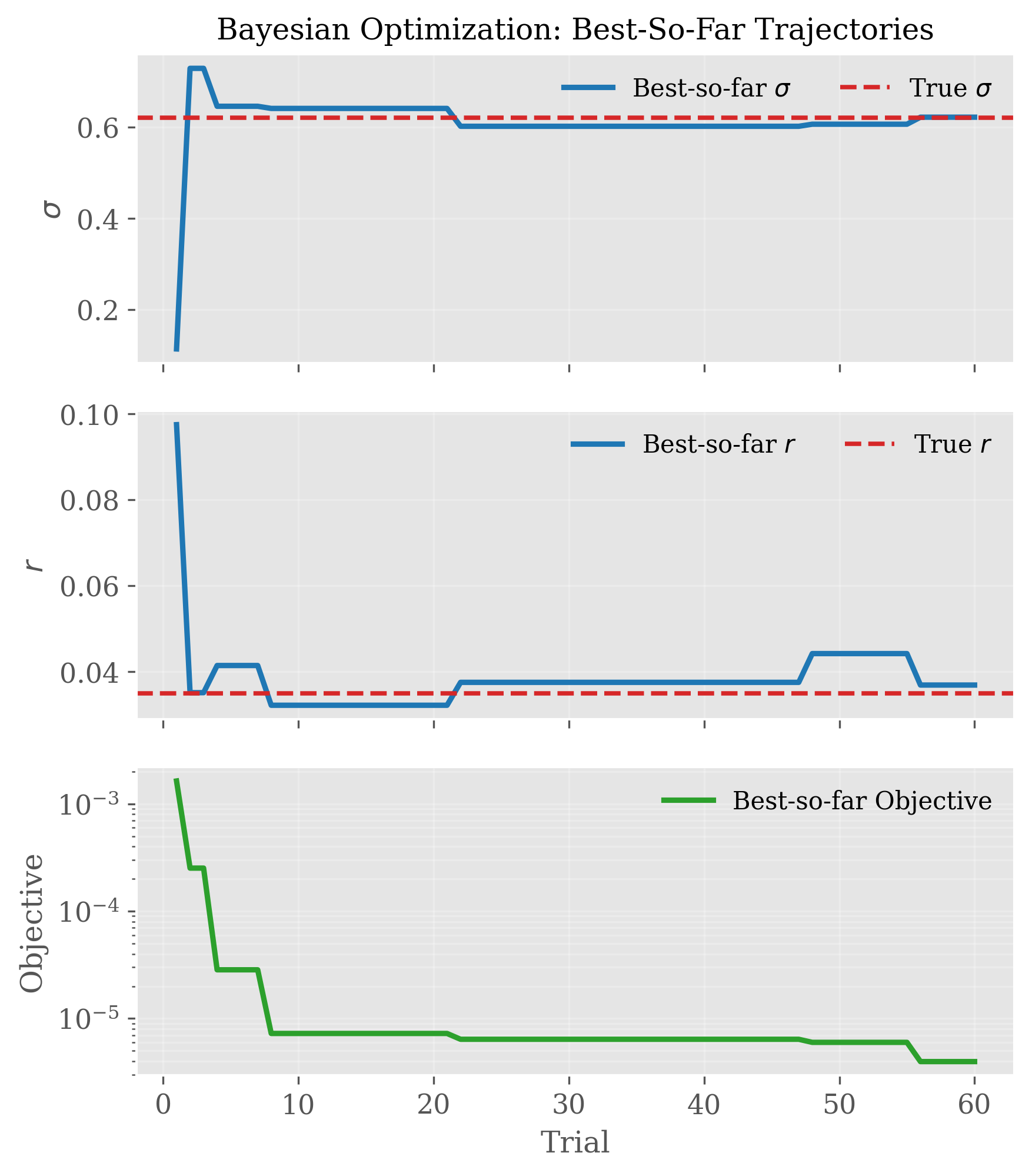}
\caption{An instance of Best-so-far trajectories during Bayesian optimization for the Black–Scholes inverse problem. Panels show (top) volatility $\sigma$ and (middle) risk-free rate $r$ best-so-far estimates (solid blue) alongside reference values (dashed red), and (bottom) best-so-far objective (green, log scale) versus BO iteration. The true parameters were $\sigma_{\text{true}} = 0.62$ and $r_{\text{true}} = 0.035$, while the Bayesian optimization recovered $\hat{\sigma} = 0.6216$ and $\hat{r} = 0.03691$. The objective is the mean-squared error between model predictions and observations at selected $(S,t)$ points.}

\label{fig:inverse}
\end{figure}

The inferred values of $\sigma$ and $r$ closely match the ground truth parameters, with relative errors below $10^{-3}$ across all tested cases. Figure~\ref{fig:inverse} illustrates the convergence of the inverse PIELM for a representative instance with $(\sigma,r)=(0.62,0.035)$, demonstrating smooth monotonic decay in the loss function and stable recovery of parameters.

These results indicate that the inverse PIELM formulation can accurately recover PDE parameters from observed option prices while providing uncertainty estimates through the Bayesian prior. This extends the PIELM framework beyond forward simulation to practical model calibration tasks.

\section{Conclusion}
\label{sec:conclusion}

This paper introduced \emph{Physics-Informed Extreme Learning Machines} (PIELMs) with $\tanh$ activations as fast, mesh-free surrogates for solving both forward and inverse problems in quantitative finance. Applied to the Black–Scholes and Heston–Hull–White (HHW) models, the proposed framework demonstrated that PIELMs can achieve accurate and stable option pricing while offering substantial runtime advantages over traditional solvers.

The methodology was further extended to an inverse setting, where a Bayesian PIELM formulation was proposed for recovering model parameters such as volatility and interest rate directly from noisy market data. By embedding the governing PDE into a probabilistic linear model, the approach jointly enforces data fidelity and physics consistency, enabling robust parameter inference without iterative backpropagation.

Future work will explore adaptive basis placement~\cite{schaback2007kernel}, hybrid global–local activations through composition of functions, and extensions to nonlinear or free-boundary PDEs such as American options via Curriculum Learning~\cite{DWIVEDI2025130924}.  

Overall, PIELMs combine the efficiency of closed-form learning with physics awareness, offering a promising direction for real-time financial modeling.

\bibliographystyle{ACM-Reference-Format}
\bibliography{sample-base}

\end{document}